\documentclass[aps, prd, twocolumn, superscriptaddress]{revtex4}

\usepackage{graphicx}
\usepackage{dcolumn}
\usepackage{amssymb}

\usepackage[dvips]{color} 

\begin{document}

\newcommand{\nab}[1]{\textcolor[rgb]{1,0,0}{#1}}

\newcommand{\conj}{^{*}}
\newcommand{\half}{\frac{1}{2}}
\newcommand{\vect}[1]{\mathbf{#1}}

\newcommand\beq{\begin{equation}}
\newcommand\eeq{\end{equation}}
\newcommand\bea{\begin{eqnarray}}
\newcommand\eea{\end{eqnarray}}
\newcommand\be{\begin{equation}}
\newcommand\ee{\end{equation}}
\newcommand\cM{{{\cal{M}}}}
\newcommand\ba{{\mathbf{a}}}
\newcommand\bb{{\mathbf{b}}}
\newcommand\bff{{\mathbf{f}}}
\newcommand\bx{{\mathbf{x}}}
\newcommand\bxd{\dot{\mathbf{x}}}
\newcommand\bX{{\mathbf{X}}}
\newcommand\bXA{{\mathbf{X}^A}}
\newcommand\bXB{{\mathbf{X^B}}}
\newcommand\bXd{\dot{\mathbf{X}}}
\newcommand\bzero{{\mathbf{0}}}
\newcommand\pa{\partial}
\newcommand\nn{\nonumber}
\newcommand\cd{\cdot}
\newcommand\al{\alpha}
\newcommand\ga{\gamma}
\newcommand\gau{\gamma_u}
\newcommand\de{\delta}
\newcommand\ep{\epsilon}
\renewcommand\th{\theta}
\newcommand\si{\sigma}
\newcommand\ta{\tau}
\newcommand\lra{\leftrightarrow}
\newcommand{\gsim}{\raise.3ex\hbox{$>$\kern-.75em\lower1ex\hbox{$\sim$}}}
\newcommand{\lsim}{\raise.3ex\hbox{$<$\kern-.75em\lower1ex\hbox{$\sim$}}}
\newcommand\eq{&\!\!\!=\!\!\!&}
\renewcommand\th{\theta}
\newcommand\rh{\rho}
\renewcommand\rhd{\dot\rho}
\def\d{{\rm d}}

\newcommand{\figname}[1]{\textit{#1}}

\title{On the stability of Cosmic String Y-junctions}

\newcommand{\addressIC}{Theoretical Physics, Blackett Laboratory, Imperial College, Prince Consort Road, London, SW7 2BZ, United Kingdom}
\newcommand{\addressNott}{School of Physics \& Astronomy, University of Nottingham, University Park, Nottingham, NG7 2RD, United Kingdom}
\newcommand{\addressAPC}{APC, University Paris 7, 10, Rue Alice Domon et L\'eonie Duquet, 75205 Paris Cedex 13, France}

\newcommand{\addressPYM}{Glaizer Group, 32 rue Guy Moquet, 92240 Malakoff, France}

\author{Neil Bevis} 
\email{n.bevis@imperial.ac.uk}
\affiliation{\addressIC}

\author{Edmund J.~Copeland}
\email{ed.copeland@nottingham.ac.uk}
\affiliation{\addressNott}

\author{Pierre-Yves Martin}
\email{pym.aldebaran@gmail.com}
\affiliation{\addressAPC} \affiliation{\addressPYM}

\author{Gustavo Niz}
\email{gustavo.niz@nottingham.ac.uk}
\affiliation{\addressNott}

\author{Alkistis Pourtsidou}
\email{ppxap1@nottingham.ac.uk}
\affiliation{\addressNott}

\author{Paul M. Saffin}
\email{paul.saffin@nottingham.ac.uk}
\affiliation{\addressNott}

\author{D.~A.~Steer}
\email{steer@apc.univ-paris7.fr} \affiliation{\addressAPC}

\date{\today}

\begin{abstract}
We study the evolution of non-periodic cosmic string loops containing
Y-junctions, such as may form during the evolution of a network of
$(p,q)$ cosmic superstrings. We set up and solve the Nambu-Goto
equations of motion for a loop with junctions, focusing attention on a
specific static and planar initial loop configuration. After a given
time, the junctions collide and the Nambu-Goto description breaks
down. We also study the same loop
configuration in a 
U(1)$\times$U(1) field theory model that allows composite
vortices with corresponding Y-junctions.  We show that the field
theory and Nambu-Goto evolution are remarkably similar until the
collision time. However, in the field theory evolution a new
phenomenon occurs: the composite vortices can unzip, producing in the
process new Y-junctions, whose separation may grow significantly,
destabilizing the configuration. In particular,
an initial loop with two Y-junctions may evolve to a
configuration with six Y-junctions (all distant from each other). Setting up this new configuration as an initial condition for Nambu Goto strings, we solve for its evolution and establish conditions under which it is stable to the decay mode seen in the field theory case. Remarkably, the condition closely matches that seen in the field theory simulations, and is expressed in terms of simple parameters of the Nambu-Goto system.  This implies that there is an easy way to understand the instability in terms of which region of parameter space leads to stable or unstable unzippings. 

\end{abstract}

\maketitle


\section{Introduction} \label{sec:intro}

Cosmic strings have long been known to arise naturally in a wide class of field theories \cite{Kibble:1976sj,Hindmarsh:1994re,Vilenkin:2000,Jeannerot:2003qv}, and recent work has indicated that they may also arise in superstring/M-theory \cite{Sarangi:2002yt, Jones:2003da}, for example through models of brane inflation \cite{Dvali:1998pa,Kachru:2003sx,Burgess:2004kv}. Back in the 1980's fundamental strings and cosmic strings were considered as separate. Observational constraints on the allowed cosmic string tension had to be violated by fundamental strings in order for them to unify the gauge couplings at high energy \cite{Witten:1985fp}. Moreover, whereas cosmic strings could be of cosmological scales, fundamental type I and heterotic strings were known to be unstable and their natural size was of order the Planck scale \cite{Witten:1985fp}. 

Things changed, however, with the second string revolution in which
the importance of higher dimensional brane solutions was understood,
including new types of strings, and it has since been realized that
the extra spatial dimensions could be large using warped
compactifications (for reviews see \cite{Polchinski:1998rr,Becker:2007zj}). As a consequence, it turned out that the effective
string tension of the fundamental strings could be reduced, and
depending on the compactification scheme it became possible to have
meta-stable fundamental strings of cosmological length
\cite{Copeland:2003bj,Leblond,Dvali:2003zj}.  

This has opened up the fascinating possibility that this new class of
cosmic strings, formed after a period of inflation, could provide a
unique window on string theory, via measurements of their imprint on
the cosmic microwave background radiation
\cite{Albrecht:1997nt,Contaldi:1998mx, Bevis:2006mj,
  Bevis:2007gh,Fraisse:2007nu, Pogosian:2008am}, their lensing of
distant galaxies \cite{Sazhin:2006fe, Shlaer:2005ry,Kuijken:2007ma} or
through their production of gravitational waves
\cite{Damour:2000wa,Siemens:2006yp,Hogan:2006we} and massive particles
\cite{Vincent:1997cx, Hindmarsh:2008dw}. 

When two cosmic strings intersect, the traditional lore dictates that they intercommute (swap partners), which in the case of a string looping back and intersecting itself results in the loop being cut off. For traditional field theory strings (essentially Abelian Higgs strings close to the Bogolmonyi limit)
this intercommutation occurs with effectively unit probability at each intersection
\cite{Shellard:1987bv,Polchinski:1988cn,Copeland:1986ng,Shellard:1988zx,Matzner:1988}. However
in the cosmic superstring case the intercommutation probability would
be reduced due to the presence of extra dimensions
and a small string coupling constant \cite{Jackson:2004zg,Jackson:2007hn,Hanany:2005bc,Avgoustidis:2005nv}.  

Additionally, one of the properties of cosmic superstrings is that there can be two different types of string, along with stable composites of the two. That is, in addition to fundamental (F) strings, there are Dirchelet (D) strings and also ($p$,$q$) composites of $p$ F-strings and $q$ D-strings. Where composites split up into more basic elements, there will be Y-shaped junctions, something not present in the cosmic string scenario that is conventionally considered. This can lead to very different dynamics when strings collide and a crucial outstanding question is how these would affect the properties of the string network and what the observational consequences would be?

However, this question is not restricted to cosmic superstrings. Even the simplest model of gauge strings (the Abelian Higgs model) can provide stable composites if the gauge coupling is set sufficiently high. Indeed, for part of this article we will use the double U(1) model \cite{Saffin:2005cs} to explore the properties of the Y-junctions that it permits, thereby increasing the applicability of our investigations, although it should be noted that the mass spectrum of the bound states is a specific prediction for cosmic superstrings that is not reproduced by these field theoretic models. 

The focus of this article is simple, a detailed comparison of the two approaches that have been adopted to describe the dynamics of strings containing Y-junctions. The first, developed by Copeland, Steer and Kibble in  \cite{Copeland:2006eh,Copeland:2006if} (CKS),  is based on modifying the Nambu-Goto action to include junctions, and the second is to describe the strings as composite objects in terms of an underlying classical field theory that will allow for the formation of junctions \cite{Saffin:2005cs},\cite{Spergel:1996ai}-\cite{Bevis:2008hg}. Such a comparison has a well established history for ordinary cosmic strings, where it has been shown that the Nambu-Goto action for a relativistic 1D string is an excellent approximation to the dynamics of widely separated gauge cosmic strings as long as the  curvature is on scales far greater than the microscopic string width. Once that regime breaks down, the Nambu-Goto approximation is no longer such a good approximation to the full field theory (for a review see \cite{Hindmarsh:1994re,Vilenkin:2000}). How does the comparison fare for these more complicated arrangements representing strings with junctions, and full loop configurations? Recently Bevis and Saffin \cite{Bevis:2008hg} showed that the results obtained by CKS for the collision of straight strings are consistent with those found using field theory simulations of the double U(1) model, once the composite region grows to be much larger than the string width. This is quite remarkable given the fact that the conditions for Nambu-Goto applicability are breeched in the region of the Y-junction at all times (a related result was also shown for a different field theory model in \cite{Salmi:2007ah}). 

In the following, we employ the CKS approach in numerical simulations for the first time, exploring scenarios not available through analytical means and in doing so we provide a quite different comparison with field theory simulations to that previously performed. Of particular note, we are able to address directly questions such as how closely the two approaches match, where the deviations arise and whether we can safely adopt the Nambu-Goto formalism to describe the evolution of loops containing Y-junctions. The answers to these questions, and in particular the last question is important. It is far easier to evolve a loop based on a modified Nambu-Goto action than on a full field theory action, hence it is possible to analyse a broader class of configurations and obtain more believable statistics concerning the distribution of cusps and kinks on these loops -- which in turn is important if we want to establish gravitational wave templates for these objects. Motivated by an instability we see in our field theory simulations, we 
also study it in the context of the Nambu-Goto approach, both analytically and numerically. This then allows us to establish the criteria in terms of a simple angular parameter, under which the junctions are stable to unzipping. This in turn will be important in the future when modeling a network of strings with junctions numerically using the Nambu-Goto equations. 

\begin{figure}
\resizebox{\columnwidth}{!}{\includegraphics{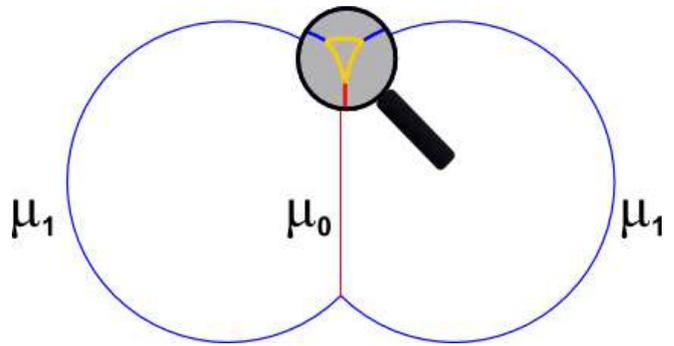}}
\caption{An example of a loop configuration with multiple junctions: the butterfly configuration with a central string of tension $\mu_{0}$ and two arc strings of tension $\mu_{1}$. The basic butterfly configuration has just two junctions, but we find that under certain situations these can decompose, as indicated by the magnified region, with a single junction splitting into three junctions that then continue to separate.}
\label{fig:setup}
\end{figure}

For simplicity, and to allow a direct comparison, we concentrate in this paper on a particular type of loop  configuration, shown in Fig.~\ref{fig:setup}. It resembles a butterfly and looks somewhat like the result of two co-planar loops colliding. We stress that this is not intended to represent a likely cosmological scenario, but presents an interesting case with which we may explore the properties of Y-junctions, including a new feature, their stability to decomposition into three separate Y-junctions.

In section \ref{sec:nambugoto}, we discuss the Nambu-Goto approach and describe our numerical technique, for loops with just two junctions and also loops with multiple junctions.  This is followed in section \ref{sec:fieldtheory} with a description of the field theoretic simulations, and then in section \ref{sec:results} we present our results for each of the approaches including the identification of an instability in the field theory case that is then seen for the first time to also be present in the Nambu-Goto case, once suitable initial conditions are chosen. We conclude in section \ref{sec:discuss} and finish with three appendices giving detailed derivations of some of the results used in the main paper. 


\section{Nambu-Goto simulations} \label{sec:nambugoto}

\subsection{Equations of motion} \label{sec:nambugotoA}

In this section we set up the Nambu-Goto equations of motion for a string loop with $J$ junctions, generalising the aforementioned CKS approach that studied the dynamics of one junction of three strings. 
For example, a loop with a few junctions is shown in Fig.~\ref{fig:setup}, where the colours differentiate the different charges and tensions of the strings.  

\subsubsection{Case of two junctions}

For simplicity, we first derive the equations of motion for a loop with
two junctions, which we label by the index $J=(A,B)$,
and three strings. We parameterize the position of
the $i$th string ($i=1,2,3$) as  
\be
x_i^\mu(\tau,\sigma_i),
\ee
where $\tau$ and $\sigma_i$ are the world-sheet coordinates (note that $\tau$ is chosen to be the same for all three strings). The induced metric on the world-sheet for string $i$ is 
\be
\gamma_{ab}^i = \frac{\partial x^\mu_i}{\partial \sigma^a}   \frac{\partial x^\nu_i}{\partial \sigma^b}    \eta_{\mu \nu},
\ee
where $a=(\tau,\sigma_i)$ and $\eta_{\mu \nu}$ is the 4-dimensional Minkowski metric.  Below, a dot/dash denotes a derivative with respect to $\tau$/$\sigma_i$ respectively. 
The values of the world-sheet coordinate $\sigma$ at the junction are denoted by $s^J_i$ and are generally $\tau$-dependent.
We are free to choose the direction of increasing $\sigma$ on each string with respect to a given junction. 
In the case of two junctions, we may take $\sigma$ to increase from junction $A$ to junction $B$ so that
\be
s^A_i(\tau) \leq \sigma_i \leq s^B_i(\tau).
\ee
For multiple junctions, as discussed below, this choice cannot always be made.
The positions of the junctions are
\be
X^\mu_J(\tau) = x_i^\mu(\tau,s_i^J(\tau)) \qquad {\rm for \; all \; } i.
\label{con}
\ee

During the evolution of the loop, it may occur that for one of the strings and at a certain time $\tau_c$, $s^A_i(\tau_c)=s^B_i(\tau_c)$.  In that case, the physical length of the string between the junctions vanishes, and the junctions collide.  
More generally, junctions $A$ and $B$ collide when $X^\mu_{A}(\tau_c) = X^\mu_{B}(\tau_c)$ which may occur even if $s^A_i(\tau_c) \neq s^B_i(\tau_c)$ and the length in string $i$ does not vanish.  Since in string-theory the outcome of the collision of junctions is not well understood, we will stop our NG simulations of loops at the collision time. The field-theory simulations presented in Sec.~\ref{sec:fieldtheory} can, of course, go beyond this time.

In the absence of background fluxes and after dilaton stabilisation, the dynamics of a {\it single infinite} $(p,q)$-string in flat spacetime is given by the Dirac Born Infeld (DBI) action \cite{Copeland:2007nv}
\be
S_{\rm DBI} = -\bar{\mu} \int  \d \tau  {\rm d} \sigma \sqrt{-|\gamma_{ab}+ \lambda F_{ab}|},
\label{DBI}
\ee
where $\bar\mu=|q|/(g_s \lambda)$ is the tension of $q$ coincident D-strings, $\lambda = 2\pi \alpha'$, with $ \alpha'$ the Regge-slope parameter, and $g_s$ is the perturbative string coupling. $F_{ab}$ is the electromagnetic tensor on the string world-sheet, 
and the electric flux density is the momentum conjugate to the electric field $p=\partial L_{\rm DBI}/\partial F_{\tau \sigma}$.
The dynamics of {\it three semi-infinite} $(p,q)$-strings meeting at a
junction was discussed in \cite{Copeland:2007nv} where it was shown that the
resulting equations of motion are exactly equivalent to those obtained
by using the Nambu-Goto action for each string, {\it provided} the
$i$th string tension in the Nambu-Goto action is taken to be given by 
\be
\mu_i = \sqrt{p_i^2 + \left(\frac{q_i}{ g_s} \right)^2},
\label{tensionF}
\ee
and one imposes charge conservation at the junction
\be
\sum_i p_i=0 \qquad \sum_i q_i=0.
\label{chargeconsv}
\ee
For this reason, and for simplicity, we assume that the dynamics of each individual segment of string is determined by the Nambu-Goto action.  

In the conformal gauge
\bea
\label{conformal}
\gamma^i_{\tau \tau} + \gamma^i_{{\sigma_i} {\sigma_i}} = 0 \text{ ; } \qquad 
\gamma^i_{\tau {\sigma_i}} = 0,
\eea
the Nambu-Goto action for the three strings of tensions $\mu_i$ joined by two junctions is 
 \bea 
 S&=&-\sum_i\mu_i \int \d\tau \int \d\si_i\, \big[\Theta(s^{B}_i(\tau)-\si_i)\Theta(-s^A_i(\tau)+\si_i) 
 \nonumber
 \\
 && \qquad \qquad \qquad \qquad \qquad
 \times \sqrt{-x_i'^2 \, \dot x_{i}^{2} } \big]
 \nonumber\\
 &&+\sum_{J=(A,B)}\sum_i\int \d \tau \,
 f^J_{i  \mu}  \cdot [x_{i}^\mu( \tau,s^J_i (\tau) )-{X}_J^\mu(\tau)], 
 \label{action1} 
 \eea
where $\mu_i$ is given in Eq.~(\ref{tensionF}) and the four-vector Lagrange multipliers $f^J_{i  \mu}(\tau)$ impose the constraints given in Eq.~(\ref{con}).

Varying the action (\ref{action1}) with respect to $x^\mu_{i}$ yields the usual equation of motion for a string in Minkowski space-time (away from the junction), namely the wave equation
\begin{equation}
\label{eq:wave}
\ddot{x}_i^{\mu} -{x_{i}^{\mu} }''=0 \; \; \Longrightarrow \; \;
x_{i}^{\mu}  = \frac{1}{2} \left[ a^\mu_i (u_i) +  b^\mu_i(v_i) \right],
\end{equation}
where 
\be
u_i = \sigma_i + \tau \text{ ; } \qquad v_i = \sigma_i - \tau.
\ee
From the conformal gauge conditions (\ref{conformal}) the ``left" and ``right" movers satisfy
\be
a_i'^2 = 0 \; , \qquad b_i'^2 = 0.
\label{confg}
\ee
Furthermore, varying the action with respect to $X^\mu_J$, imposing
the temporal gauge and using the boundary conditions gives energy
conservation equation at each junction (see \cite{Copeland:2006eh} and
appendix \ref{sec:appendixA} for details): 
\begin{equation}
\label{eq:energycks}
 \mu_1\dot{s}^J_1+ \mu_2\dot{s}^J_2+ \mu_3\dot{s}^J_3=0.
\end{equation}
As the junction is moving, some of the strings will have $\dot{s}^J_i>0$ while others $\dot{s}^J_i<0$. These represent growing/shrinking of the string not only in $\sigma$-space, but also in real space. The rate of creation of one string must balance the disappearance of other(s). 

Given an arbitrary initial loop configuration ($\bx_i(0,\si_i)$, $\dot{\bx}_i(0,\si_i)$), we aim to solve the above equations in time in order to give the full loop evolution and hence ($\bx_i(t,\si_i)$, $\dot{\bx}_i(t,\si_i)$). We will now concentrate on junction $B$. Let us first define the ``communication route" on the loop for a time $t$ at a junction $B$ by  
\be
\ell^B_{\rm min}(t) = \mathop{\rm min}_{i} |{ s_i^B(t)-s_i^A(0)|}.
\label{chara}
\ee
For a time $t < \ell^B_{\rm min}$ junction $B$ receives no influence from the dynamics at junction $A$, with the initial
conditions specifying the incoming waves $b^\mu_i$ at $B$, so determining the behaviour of the junction. For times beyond $\ell^B_{\rm min}$, on the other hand, vertex $B$ is affected by the dynamics at vertex $A$ and the problem becomes more complex. While in Appendix \ref{sec:appendixB} we shall obtain analytical results that are valid for all times before the two junctions meet, this should be considered a special case and
in general numerical methods are required.

The general formalism is fully described in appendix \ref{sec:appendixA}. The amplitudes of the outgoing waves $a_{i}'^{\mu}$ are determined by the amplitudes of the incoming waves $b_{i}'^{\mu}$ at the junction, and we find
\bea
( \dot s^B_{i} + 1)  a_{i}'^{\mu} = b_{i}'^{\mu}
(1-\dot{s}^B_i)-\frac{2}{\sum_k \mu_k} \sum_j{\mu}_j  (1-\dot{s}^B_j)
b_{j}'^{\mu}, \nonumber \\
\label{thus}
\eea
while the evolution of $\dot{s}^B_i$ is determined by
 \beq\label{sdot_b}
1-\dot s^B_i(t)=
 \frac{(\sum_j\mu_j) M_i(1-c^B_{i}(t))}{\mu_i \sum_k M_k(1-c^B_{k}(t))},
 \label{sdot}
 \eeq
where 
 \bea c^B_{1}(t)&=&\bb'_2(v^B_{2}(t))\cd\bb'_3(v^B_3(t))  ,
 \label{cB}
 \\
M_1 &=&  \mu_1^2-(\mu_2-\mu_3)^2,
\eea
and cyclic permutations.  Note that it follows from (\ref{sdot}) and imposing causality that one must have $M_j \geq 0$. This is automatically satisfied by $(p,q)$ strings with tensions satisfying Eqs.~(\ref{tensionF}) and (\ref{chargeconsv}). At vertex $A$ the procedure is similar, though the incoming waves are now given by the $a_{i}'^{\mu}$.

\subsubsection{Multiple junctions}

Finally, we now generalize the above formalism to a string loop with
more than two junctions. The important point here is that it is no
longer possible to choose the $\sigma_i$'s all to increase (or all to
decrease) into a given junction.   
This can be seen, for example, by considering carefully the triangular section of the loop shown in figure 1.  Hence we
must treat cases in which there is a different orientation between the
three strings at a junction. 

To do so we associate a further parameter $\delta_i^J$ with the 3
strings meeting at junction $J$. If $\delta^J_i = +1$ then on string
$i$, $\sigma_i$ increases into the junction. If, on the other hand,
$\delta_i = -1$ for string $i$ then $\sigma_i$ decreases into the
junction. In the action, $\delta^J_i$ therefore ensures invariance 
under $\sigma$-reparametrisations, and full details are given in
appendix \ref{sec:appendixA}.  Note also that the values of 
$\delta^J_i$ for two junctions that are connected by the same piece
of string are related.
The energy conservation equation at the junction is now generalized
(see appendix \ref{sec:appendixA}) to  
\begin{equation}
\label{eq:energy}
 \delta^J_1\mu_1\dot{s}^J_1+ \delta^J_2\mu_2\dot{s}^J_2+ \delta^J_3\mu_3\dot{s}^J_3=0.
\end{equation}
For example, consider $ \delta^J_1 =  \delta^J_2 = +1$ but changing the $\sigma$ orientation of string $3$.  Then $\dot{s}^J_3$ picks up a minus sign but $\delta^J_3 = -1$, so the energy equation is unchanged and energy is still conserved. 
Finally, the formula for the $\dot{s}_i$ as a function of the incoming waves at a given junction $J$ is
\beq
\dot s^J_i(t) = \delta_i \left(1-
 \frac{(\sum_j \mu_j) M_i(1-c^J_{i}(t))}{\mu_i \sum_k M_k(1-c^J_{k}(t))}\right),
 \label{sdotgen}
 \eeq
where 
 \bea c^J_{i}(t)&=& {\bf Y}_j \cd {\bf Y}_k,
 \label{cBgen}
\eea
with $i\neq j \neq k$ and
\[ {\bf Y}_j = \left\{ \begin{array}{lll}
          \bb'_j& & \mbox{if \ $\delta^J_j = +1$}\\
        -\ba'_j& & \mbox{if \ $\delta^J_j = -1$}.\end{array} \right. \]

\subsection{Numerical approach}

While the above equations allow for analytical calculations for some particularly simple initial conditions (typically semi-infinite straight strings) \cite{Copeland:2006eh,Copeland:2006if}, here we consider more complex loop configurations for which numerical methods are necessary. 
We proceed in the following way.  For every string $i$ connecting two junctions,
we work entirely with $\ba'$ and $\bb'$, reconstructing the closed string position $\bx(t,\si_i)$ and velocity $\dot{\bx}(t,\si_i)$ only when necessary. The initial conditions fix the initial string tensions $\mu_i$ satisfying the triangle inequalities $M_j \geq 0$, as well as $\ba'_i(\sigma_i)$ and $\bb'_i(\si_i)$ between all the junctions.
Then we calculate the $c_i^J(t=0)$ from which $\dot{s}_i^J(t=0)$ is determined using Eq.~(\ref{sdotgen}). Then at time $\delta t$,  $s_i^J(\delta t)$, $u_i^J(\delta t)$ and $v_i^J(\delta t)$ can be calculated.  The last step is to use (\ref{thus}) (or the appropriate generalized formula in the case of different relative orientation of the strings, see for example eq.~(\ref{eq:thusgen})) 
at each junction to extend the domain of definition of $\ba_i'(u)$ and $\bb_i'(v)$.  The time loop then continues.

Our simulation ends whenever the length of one string goes to zero (so whenever two junctions meet).
However, the field theory simulations described in the next section can of course continue beyond this time.

\subsection{Initial conditions} \label{sec:butterfly}

We consider below {\it two} different initial conditions that, as we shall describe, are closely related. 

The {\it first} is the butterfly configuration shown in
Fig.~\ref{fig:setup}, consisting of only two junctions (in other
words, the small triangle shown in the magnifying glass is not present
in this initial configuration). More specifically, this planar
configuration consists of a straight string with tension $\mu_0$ and
two circular arcs with equal tensions $\mu_1$. We take the
strings to be initially static at all points: at the site of the
Y-junctions this requires $\dot{s}_{j}=0$ which is satisfied when the
vector sum of tensions at the junction is zero, namely

\begin{equation}
\sum_j \mu_j \frac{\mathbf{x}_j'}{|\mathbf{x}_j'|} 
= 0.
\label{vec-sum}
\end{equation} 
This corresponds, physically, to the fact that there can be no change
in momentum occurring within a small volume surrounding the junction
if the situation is static, and is in fact the same condition as derived in Appendix A, Eq.~(\ref{Xeqn}) for the case of the static string. This condition yields a relationship
between the string tensions, the radius $r$ of the two arcs making the
butterfly ``wings'', and the distance $x$ of straight string from their centres:

\begin{equation}\label{eq:Rdef}
\frac{\mu_0}{2\mu_1} = \frac{x}{r} \equiv R.
\end{equation}

The {\it second} initial condition is  a modification of the first  initial condition, in that we add, at each junction, the triangular shape shown in the
magnifying glass in Fig.~\ref{fig:setup}.  Hence this initial
configuration consists of $9$ strings and $6$ junctions, all of which are static.  All the
strings within the small triangles are taken to be arcs of circles,
and are all given the same tension $\mu_2$. The reason for
considering this second initial condition is that our field theory
simulations of the simple butterfly configuration (with initially only
two junctions) will show (see section \ref{section:stab}) that the
central straight string can be dynamically unstable into splitting into a
configuration very much like the one shown in figure 1.  This
dynamical splitting cannot be accounted for in the Nambu-Goto equations, and
hence we are forced to put it in by hand as an initial perturbation.
For that reason the initial size of this perturbation is given by a
free parameter $h$, (effectively the distance between one of the old and newly created junctions), which does not affect the general physical
behaviour if initially small; this only depends on the relative
tensions of the strings. The fact that the string is initially static
sets all angles, shown in Fig.~\ref{fig:setup}, to be functions of the
tensions $\mu_i$ and the parameter $h$.   

Finally, we should
note that the use of circular segments to describe the small perturbation is
somewhat arbitrary, and it restricts the range of allowed angles (or
effectively the value of the tensions $\mu_2$ in the perturbation) that can be used to construct this
second static initial configuration. However, we  
believe that this choice does not have any dramatic effect on the
local dynamics, since it is only the local curvature at $t=0$ around a
given junction that really affects its evolution. We will illustrate
this in section \ref{section:stab}, when studying the stability of
Y-junctions. 


\section{Field theoretic approach} \label{sec:fieldtheory}

\subsection{The field theory model} \label{sec:fieldtheoryA}

While the Nambu-Goto formalism is relatively easy to analyse numerically, it does not necessarily give a complete description of Y-junctions --- for instance, one might expect important interactions between the strings close to and at the junctions, and these are not included in the Nambu-Goto action.
For that reason we also study the butterfly  configuration, listed
above, using a field theoretic model of gauge strings that permits the
formation of junctions.

Strings with junctions can be formed in numerous different symmetry breaking schemes: here we study bound states and corresponding Y-junctions of the double U(1) model \cite{Saffin:2005cs}, which has Lagrangian density:
\begin{eqnarray}
\mathcal{L} 
& = & 
-\frac{1}{4}F_{\mu\nu}F^{\mu\nu} 
-(D_{\mu}\phi)\conj (D^{\mu}\phi)
-\frac{\lambda_{1}}{4}\left(|\phi|^{2} - \eta^{2} \right)^{2}\nonumber\\
& &
-\frac{1}{4}\mathcal{F}_{\mu\nu}\mathcal{F}^{\mu\nu} \!
-(\mathcal{D}_{\mu}\psi)\conj (\mathcal{D}^{\mu}\psi)
-\frac{\lambda_{2}}{4}\left(|\psi|^{2} - \nu^{2} \right)^{2}\nonumber\\
& &
+ \kappa  \left( \left|\phi\right|^2 - \eta^{2} \right)\left( \left|\psi\right|^2 - \nu^{2} \right).
\end{eqnarray}
The notation used follows Refs. \cite{Saffin:2005cs, Bevis:2008hg} in that $\phi$ and $\psi$ are complex scalar fields, each belonging to one of the two Abelian Higgs models that are coupled together via the final term. The gauge-covariant derivatives in the two halves of the model are:
\begin{eqnarray}
D_{\mu}\phi & = & \partial_{\mu}\phi - ie A_{\mu}\phi, \\ 
\mathcal{D}_{\mu}\psi & = & \partial_{\mu}\psi - ig B_{\mu}\psi, 
\end{eqnarray}
and the anti-symmetric field strength tensors are:
\begin{eqnarray}
F_{\mu\nu} & = & \partial_{\mu}A_{\nu} - \partial_{\nu}A_{\mu}, \\
\mathcal{F}_{\mu\nu} & = & \partial_{\mu}B_{\nu} - \partial_{\nu}B_{\mu},
\end{eqnarray}
for gauge fields $A_{\mu}$ and $B_{\mu}$. Finally, $\eta$ and $\nu$ are constants that set the energy-scales of the two halves of the model and {\it e.g.~}$\lambda_{i}$ and $\kappa$ are dimensionless coupling constants.

For $\kappa=0$ the two halves are uncoupled and the topologies of the
vacuum manifolds, each obeying a local U(1) symmetry, require that
local string solutions are present \cite{Kibble:1976sj} (see
Refs. \cite{Vilenkin:2000,Hindmarsh:1994re} for reviews), and take the
form of the Nielsen-Olesen vertex \cite{Nielsen:1973cs}. For the
$\phi$-half of the model, as a loop enclosing the string is traversed
in space, with a net non-zero winding of the phase of $\phi$ equal to $2\pi p$
($p \in \mathbb{Z}$), then $|\phi|$ is forced to zero along the string
in order for the field to be smooth. As a result the field does not
lie on the vacuum manifold $|\phi|=\eta$ near the core and the string
carries significant potential energy. It additionally carries a
quantized amount of pseudo-magnetic flux, equal to $2\pi p/e$ in the
case of a $\phi$ string, since the gauge field acts to reduce the
phase gradients and so acquires a significant curl close to the
string. 

For $\kappa$ in the range $0<\kappa<\half\sqrt{\lambda_{1}\lambda_{2}}$, two parallel strings from each half can coalesce to reduce the potential energy and hence there are bound string states in this model \cite{Saffin:2005cs}. Note, however, that for $\kappa > \half\sqrt{\lambda_{1}\lambda_{2}}$ the potential term is unbounded from below and the model hence becomes unphysical.

\subsection{Collision of straight strings}

The intersection of straight strings has been studied in the above model \cite{Bevis:2008hg}. It was found that composite states do indeed form and that their growth rates at late times are well predicted by the Nambu-Goto formalism. However, the Nambu-Goto solution differs from the field theory case as the moment of collision approaches, (see also
\cite{Salmi:2007ah} for a similar analysis in the Type I regime of the
Abelian Higgs model.) This is partly because any attraction of one string to the other just prior to the intersection event
means that the strings are no longer perfectly straight \cite{Bevis:2008hg}. There are additionally other model dependent effects involved in the interaction and it is hence no surprise that the CKS approach differs from the field theory simulation in predicting precisely when the intersection results in composite formation and growth.

\subsection{Numerical approach}

The numerical approach employed for the field theory simulations is largely that of Ref.~\cite{Bevis:2008hg}, but with a very different set of initial conditions. That is, we use an extension of the Moriarty et al. approach \cite{Moriarty:1988fx} for the Abelian Higgs model with the field evolution then preserving the analogue of Gauss' law in each half of the model to machine precision. We employ a uniformly-spaced cubic lattice, with spacing $\Delta x$, on which the scalar fields are defined and then the spatial components of the gauge fields are defined on the links between sites, with the time components set to zero via a gauge choice. Energy conservation is accurate so long as the uniform timestep $\Delta t$ is sufficiently small. 

We obtain the initial conditions required for the butterfly configuration by first setting up the appropriate windings in the scalar field and then applying a period of dissipative evolution in order to relax the configuration to the minimum energy configuration. That is, during this period there is an extra term in each of the equations of motion that is proportional to the first time derivative of the corresponding field and so removes energy from the system. Additionally we fix the modulus of the scalar field in the region close to the desired centre lines since otherwise the configuration would simply contract to a point during the dissipative evolution. We apply reflective boundary conditions throughout the simulation.

\begin{figure}
\resizebox{\columnwidth}{!}{\includegraphics{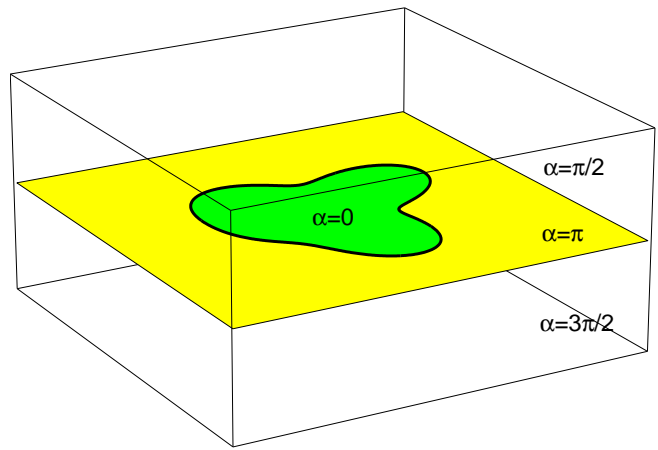}}
\caption{The initial $\phi$ phase choice for a planar (1,0) loop, ensuring a winding of $2\pi$ in the desired locations.}
\label{fig:phaseGuess}
\end{figure}

The initial choice for the phases of the $\phi$ field for a planar loop of (1,0) string is made as shown in Fig.~\ref{fig:phaseGuess}. If a site is above the plane of the loop then it is given the phase $\pi/2$, if it is below the plane then it is given $3\pi/2$, while if it is in the plane of the loop then it is given either $\pi$ if it is outside the loop or zero if it is within it. This ensures the correct winding structure of the field but it obviously yields artificially high gradients on the plane of the loop. The modulus of the scalar field is initially chosen so that the field lies on the vacuum manifold, except close to the string centre lines, as will be explained momentarily. During the dissipative evolution the gauge field, which is set to zero initially, quickly grows to counter these phases gradients, while the phase and modulus of $\phi$ rapidly adjust themselves in order to minimize the energy. Obtaining a (0,1) loop can be achieved by simply swapping $\psi$ for $\phi$ in the above argument, while a (1,1) loop is yielded by setting up the phases appropriately in both fields. Higher winding numbers cannot be achieved by the direct application of the above approach and will be discussed below.

The modulus of the scalar fields is set inside a tube around the string centre-line according to the solution for an infinite straight string. For a winding $2\pi m$ in the phase of $\phi$ and $2\pi n$ in the phase of $\psi$, this has the following form for small displacements $r$ from the string centre: 
\begin{eqnarray}
\phi(r) \approx C r^{m},\\
\psi(r) \approx D r^{n}.
\end{eqnarray}
The constants $C$ and $D$, which are dependent on the choice of $m$ and $n$ cannot be found analytically, but are solved for using essentially the approach of Ref.~\cite{Saffin:2005cs}. Note that if $m$ is finite but $n$ is zero, then even though there is no winding in $\psi$, its modulus is still less than $\nu$ near the string as this lowers the total potential term energy. However, $|\psi|$ does remain finite as  $r\rightarrow 0$. In principle we could fix $|\psi|$ close to the string in this case also, but we choose not to since it would not greatly aid the fixing of the string position and we wish to minimize the artificial restrictions enforced.

\begin{figure}
\resizebox{\columnwidth}{!}{\includegraphics{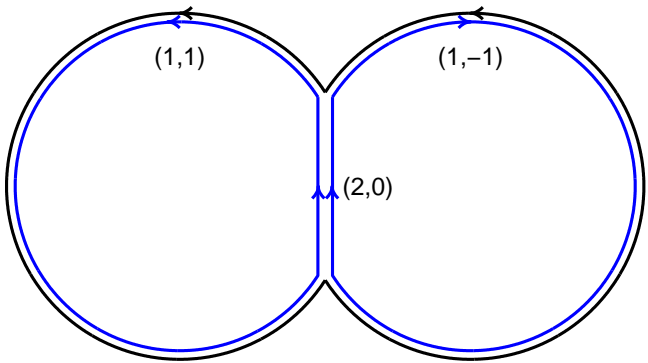}}
\caption{The fluxes present in the butterfly configuration yielded by the superposition of a (1,1) loop and a (1,-1) loop.}
\label{fig:butterflyFields}
\end{figure}

The butterfly configuration illustrated in Fig.~\ref{fig:butterflyFields} can be constructed by the superposition of a \mbox{(1,1)} loop and a \mbox{(1,-1)} loop after a period of dissipation. Since the equations of motion are non-linear there is no precise means to do this, however a good approximation is simply to sum the gauge fields from each loop $A_{\mu}^{+}$ and $A_{\mu}^{-}$ to give the total:
\begin{equation}
A_{\mu} = A_{\mu}^{+} + A_{\mu}^{-},
\end{equation}
where the $+$ and $-$ refer to each loop. 
Then for the scalar fields:
\begin{equation}
\frac{\phi}{\eta} = \frac{\phi_{+}}{\eta} \; \frac{\phi_{-}}{\eta}, 
\end{equation}
results in a superposition of complex phases \cite{Shellard:1987bv,Copeland:1986ng, Matzner:1988, Moriarty:1988fx}. Furthermore, at distances far from any string set up in $\phi_{+}$ (such that the field is approximately constant and close to its vacuum) the form of $\phi$ is essentially that found in $\phi_{-}$. Using these equations, the time derivatives must then superpose as:
\begin{eqnarray}
\partial_{t}A_{\mu} & = & \partial_{t}A_{\mu}^{+} + \partial_{t}A_{\mu}^{-},\\
\eta \; \partial_{t}\phi & = & \phi_{+}\partial_{t}\phi_{-} + \phi_{-}\partial_{t}\phi_{+}.
\end{eqnarray}
While the wings are largely unaffected by this process and remain close to the minimum energy solution, a further period of dissipation is required to relax the central region, because of the significant interference between the two loops. For the case illustrated in Fig.~\ref{fig:butterflyFields} this includes the cancellation of the fluxes in $\psi$ along the central string, which greatly reduces the energy per unit length of that segment. 


\section{Results} \label{sec:results}


We shall describe our results in units such that the initial radius of the circular arcs is unity, since this is a useful reference length scale. For the field theory simulations we set $2=\lambda_{1}=\lambda_{2}=2e^{2}=2g^{2}$, since these values are the standard choices \cite{Saffin:2005cs, Bevis:2008hg, Urrestilla:2007yw} and neatly yield the relevant values of $\kappa$ in the range $0<\kappa<1$. We have studied a number of values of $\kappa$ in this range, with $0.8$ and $0.95$ being the two main ones. We additionally set $\eta=\nu$ and hence there is complete symmetry between the two halves of the model, while we then choose the value of $\eta$ to set the ratio of the string width (roughly $1/\eta$) relative to the arc radius. Note that although the two Abelian Higgs models are each individually in the Bogomolnyi limit \cite{Bogomolny:1975de}, with finite coupling between them, the $\psi$ field affects the energy per unit length of a $(m,0)$ string even though there is no winding in $\psi$ and for example $(2,0)$ strings are stable (see table \ref{tab:mu}).

In order to compare field theoretic results with those from the Nambu-Goto simulations we additionally calculate the string tensions in the case of infinite straight strings via the method of \cite{Saffin:2005cs}. We then use these as $\mu_{0}$ and $\mu_{1}$ in the Nambu-Goto case, although of course we are also free to choose arbitrary values for these in the Nambu-Goto case. The tensions used are shown in Table \ref{tab:mu}.

\begin{table}
\begin{ruledtabular}
\begin{tabular}{ccc}
$\kappa$                     & 0.80  & 0.95 \\
\hline
$\mu_{(1,0)} / 2\pi\eta^{2}$ & 0.864 & 0.728 \\
$\mu_{(1,1)} / 2\pi\eta^{2}$ & 1.452 & 1.133 \\
$\mu_{(2,0)} / 2\pi\eta^{2}$ & 1.622 & 1.271 \\
$R[(1,0)+(0,1)\rightarrow(1,1)] $ & 0.840 & 0.778 \\ 
$R[(1,1)+(1,-1)\rightarrow(2,0)] $ & 0.559 & 0.561 \\
\end{tabular}
\end{ruledtabular}
\caption{\label{tab:mu}The energy per unit length of an infinite static string, with parameters $\lambda_{1}=\lambda_{2}=2e^{2}=2g^{2}=2$, $\eta=\nu$ and two values of $\kappa$ as indicated. Note that an Abelian Higgs string of unit winding would yield $\mu=2\pi\eta^{2}$\cite{Bogomolny:1975de}. $R$ is defined in Eq. (\ref{eq:Rdef}).}
\end{table}

For moderate values of $\kappa$ (say $0.8$) the ratio of composite to total constituent tensions, $R$, for the case of $(1,0)+(0,1)\rightarrow(1,1)$ is relatively close to unity. Even if $\kappa$ is $95\%$ of its maximum physical value then $R=0.78$ and there is only a reduction in energy per unit line of about one fifth when a $(1,0)$ string and a $(0,1)$ string combine. Further increasing $\kappa$ does not greatly reduce $R$ since most of the energy actually stems from the covariant derivative term. However a large binding energy can be achieved if the junction involves flux cancellation and therefore we also consider $(1,1)+(1,-1)\rightarrow(2,0)$. For $\kappa=0.95$ this yields $R=0.56$, which indicates that a $(2,0)$ is only slightly heavier than a $(1,1)$ string (which of course has the same tension as a $(1,-1)$ string).


\subsection{Direct comparison of field theory and Nambu-Goto strings.}

\begin{figure}
\resizebox{\columnwidth}{!}{\includegraphics{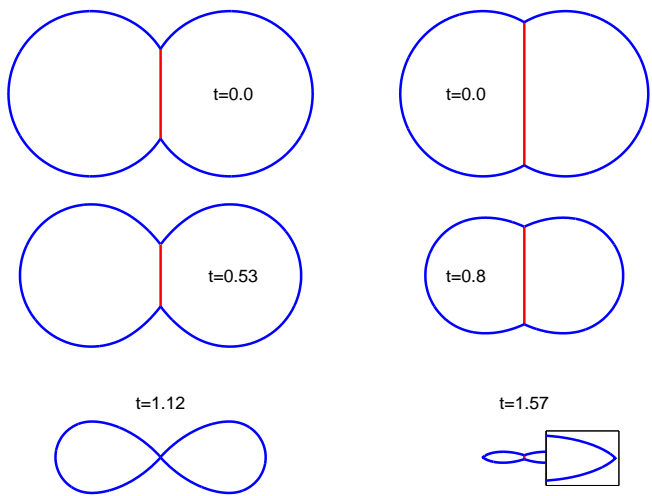}}
\caption{Results using the Nambu-Goto method with tensions set
  to match a field theory $(1,0)+(0,1)\rightarrow(1,1)$ case with
  $\kappa=0.8$ (left plot), and all tensions equal (right plot). The
  later case corresponds to $R=0.5$ and includes a magnified
  region. } 
\label{fig:NG-11+R=0.5}
\end{figure}

Before presenting our results it is useful to note that a circular Nambu-Goto
loop of unit radius collapses to a point in a time $\pi/2$. This is also the timescale for collapse of the circular arcs
in the butterfly case, though strictly this is limited to regions of
the arcs that remain causally disconnected from the junctions during
this period (see next section). For the case equivalent to $(1,0)+(0,1)\rightarrow(1,1)$ with
$\kappa=0.8$, the Nambu-Goto results shown in
Fig.~\ref{fig:NG-11+R=0.5} (left) reveal that the central bridge
string collapses on a timescale shorter than $\pi/2$. That is if
$R=0.840$ the arcs are still largely intact by the time $t=1.12$, when
the length of the central bridge string is reduced to zero. In contrast for the
case corresponding to $(1,1)+(1,-1)\rightarrow(2,0)$ with
$\kappa=0.95$, the more stable bridge remains intact until just
prior to the arc collapse: $t=1.56$ rather than $\pi/2\approx1.57$. A
very similar case to this is shown in Fig.~\ref{fig:NG-11+R=0.5}
(right), but $R$ has been set to $0.5$ instead of $0.56$. This is because
we cannot strictly continue the simulation beyond the time of bridge collapse and this
choice of $R$ yields a bridge collapse time that is just longer than the arc collapse time 
and hence the approximate moment of arc collapse can be seen in the figure.

\begin{figure}
\resizebox{\columnwidth}{!}{\includegraphics{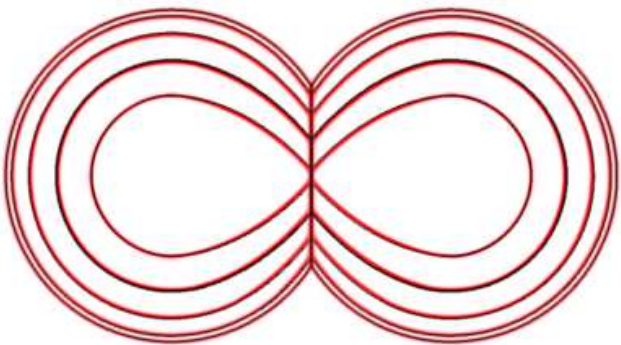}}
\caption{The evolution of the butterfly configuration
  $(1,0)+(0,1)\rightarrow(1,1)$ with $\kappa=0.8$, shown at equally
  spaced time intervals: $t=0.000$, $0.267$, $0.533$, $0.800$,
  $1.067$, with larger configurations corresponding to earlier
  times. The field theory solution is shown as a bitmap, representing
  the cumulative projection of its energy density onto the plane, 
  while the Nambu-Goto solution is shown as a solid black
  line. The field theory simulation had lattice spacing $\Delta x=0.5/\eta$, 
  with $\eta=90$. } 
\label{fig:results08}
\end{figure}

\begin{figure}
\resizebox{\columnwidth}{!}{\includegraphics{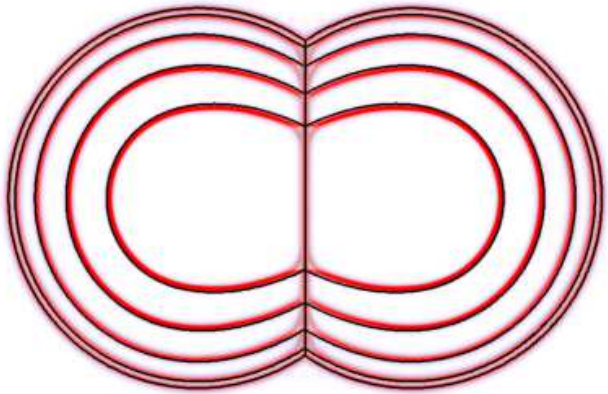}}
\caption{As in Fig.~\ref{fig:results08} but for $\kappa=0.95$ and $(1,1)+(1,-1)\rightarrow(2,0)$ .}
\label{fig:results95}
\end{figure}

In Figs. \ref{fig:results08} and \ref{fig:results95} we compare
the field theory results for these two cases with the aforementioned
Nambu-Goto ones. The snap shots shown are for equally spaced times and
hence additionally show the velocity of the strings. The primary
observation is that the agreement between the field theoretic and
Nambu-Goto results is excellent for the times shown, allowing us to extend the results from Ref.~\cite{Bevis:2008hg} to strings with global curvature rather than just straight strings with kinks, as was previously considered. 

It is also instructive to plot the physical length of the central bridge string as a function of time, as in Fig.~\ref{fig:bridgeLengthComp}.  For the Nambu-Goto case this can be found analytically via the method of Appendix~\ref{sec:appendixB} and, since this string is stationary, it is just equal to the invariant length: $s^{B}-s^{A}$. Of course, it can also be obtained from our Nambu-Goto simulations, and provides a nice test of them, however in the field theory simulations, the measurement of the string length is non-trivial.

Fortunately, the geometry of the central bridge is very simple since it merely lies along the $y$-axis at all times. Further we can detect the path of the strings by searching for lattice grid squares around which there is a net winding in the phase of one or both of the scalar fields $\phi$ and $\psi$ (using the gauge-invariant method of Ref.\ \cite{Kajantie:1998bg}). 
For the case of $(1,0)+(0,1)\rightarrow(1,1)$ we then search for net windings that lie within a certain threshold of the y-axis and calculate the bridge length as the maximum difference in $y$-coordinates. There is a small systematic error dependent upon the threshold and the angles at which the string arcs meet the bridge string, but this is not relevant. For the second case involving $(1,1)+(1,-1)\rightarrow(2,0)$, it is useful to note that, as shown earlier in Fig.~\ref{fig:butterflyFields}, there is a (0,1) string that passes around the extremes of the configuration but not along the bridge. The bridge length can hence be taken to be the distance between the two points where this string meets the $y$-axis.

Results again show excellent agreement between the Nambu-Goto and field theoretic approaches, and in fact right up to the moment of bridge collapse. However, there is a small transient departure initially in the $(1,1)+(1,-1)\rightarrow(2,0)$ case, but this is largely due to the initial conditions employed and the bridge length measurement technique. The initial conditions in the field theory case force the $\phi$ and $\psi$ string centre lines along very specific and idealized paths, but close to the junction these are not the natural paths, which are less tightly curved. Hence when the strings are released at $t=0$, they move to follow these more natural paths but overshoot slightly before being drawn back. Hence the initial disagreement in Fig.~\ref{fig:bridgeLengthComp} is just the (0,1) string attempting to follow a smoother and less kinked route across the junction and therefore moving outwards from it, but then going too far and so undergoing a few low-level oscillations. The corresponding oscillations for the other parts of the Y-junction, in which the two (1,0) strings that compose the (2,0) bridge separate slightly, can be seen in Fig.~\ref{fig:results95}. 

\begin{figure}
\resizebox{\columnwidth}{!}{\includegraphics{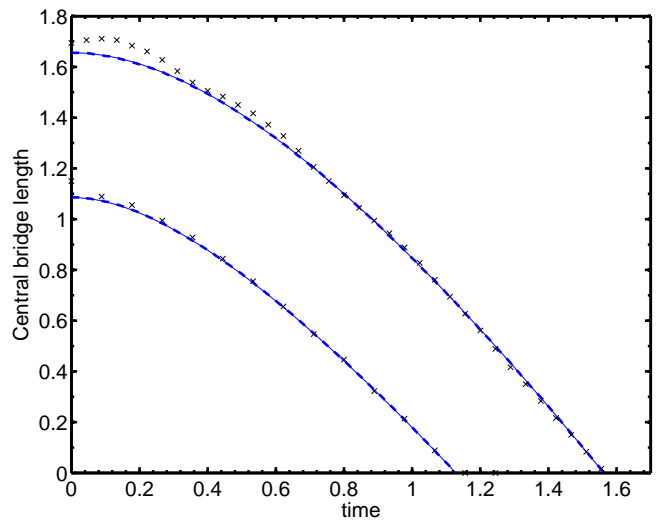}}
\caption{The length of the central bridge string as a function of time for the analytic Nambu-Goto solution (thin), the numerical Nambu-Goto results (thick, dashed) and the field theoretic results (crosses). The collection of data with lower bridge lengths is for the $(1,0)+(0,1)\rightarrow(1,1)$ case with $\kappa=0.8$ while higher bridge values correspond to $(1,1)+(1,-1)\rightarrow(2,0)$ with $\kappa=0.95$. The initial departure of the field theoretic results from the Nambu-Goto ones is due to the measurement technique and is a transient effect.}
\label{fig:bridgeLengthComp}
\end{figure}

\subsection{Collapse of circular arcs in regions causally disconnected from the junctions}

As briefly discussed above, an initially static, circular Nambu-Goto loop of unit radius, collapses to a point in Minkwoski space-time in a time $t=\pi/2$. In fact, the Nambu-Goto solution involves the loop radius undergoing simple harmonic motion with period $2\pi$. That is, the loop collapses to a point but each segment of the loop re-emerges on the other side of the loop centre. However, the apparent period of the motion is just $\pi$ because segments from one side of the loop are indistinguishable from those on the other side. Not surprisingly, the Nambu-Goto action fails to accurately describe a field theoretic model once the curvature of the string approaches the string width and hence the evolution of the two types of string differs just prior to $t=\pi/2$.

For the butterfly configuration, by the collapse time, information about the presence of the junction will have travelled around the arcs an angle $\pi/2$, which is just equal to the invariant length traversed for our chosen initial arc radius. Since our initial conditions yield an arc length of $2(\pi-\cos^{-1}(R))$ then a length $\pi-2\cos^{-1}(R)$ remains unaffected by the junctions, which is between $0$ and $\pi$ for $R$ between $0$ and $1$. Hence a fraction of the arcs will collapse to a point, reaching the speed of light for an instant, and yield a sharp kink in the string at a point in physical space. This is confirmed in our numerical simulations and can be seen in the magnified region of Fig.~\ref{fig:NG-11+R=0.5}. Note that this is not a cusp in the conventional sense, in which the string reaches the speed of light for an infinitesimal range of $\sigma$ only. In the butterfly case (with unit initial radius) the change in angle is simply equal to the invariant length involved, and so would be zero for an infinitesimal region. 

In the field theory case, the circular arcs initially follow the Nambu-Goto dynamics closely, as can be seen in Figs.~\ref{fig:results08} and \ref{fig:results95}. However, the sharp kink at the moment of collapse is not reproduced and rather than the strings emerging as an arc of "negative radius", the energy associated with the arc is dissipated away as radiation. However, the string velocity becomes highly relativistic and so the string width experiences a significant Lorentz contraction. As this happens, discretisation errors become increasingly important, hence this region is even troublesome for the field theory simulations. Since it is not the focus of this paper and merely an attribute of our idealized initial conditions, we shall not dwell on this region further.

\subsection{Stability of Y-junctions to decomposition into multiple Y-junctions}\label{section:stab}

\begin{figure}
\resizebox{\columnwidth}{!}{\includegraphics{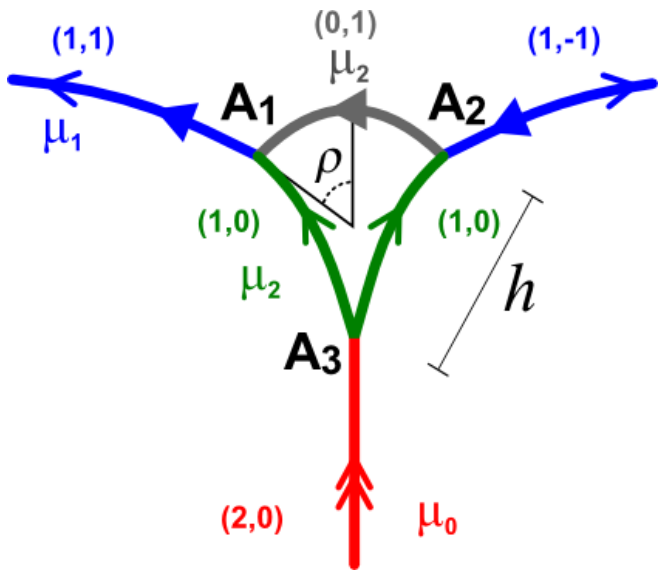}}
\caption{In the field theory, this diagram shows the possible
  decomposition of a \mbox{$(1,1)+(1,-1)\rightarrow(2,0)$} junction
  into three separate Y-junctions. For the Nambu-Goto simulations this can
  be thought of as the second initial condition (magnified region of the
  butterfly 
  configuration shown in Fig.~\ref{fig:setup}), in which the
  Y-junction has split into three Y-junctions. In this case, the
  angles shown would be determined by the tensions, $h$, and the fact
  that the junctions are initially static. } 
\label{fig:setup2}
\end{figure}

\begin{figure}
\resizebox{\columnwidth}{!}{\includegraphics{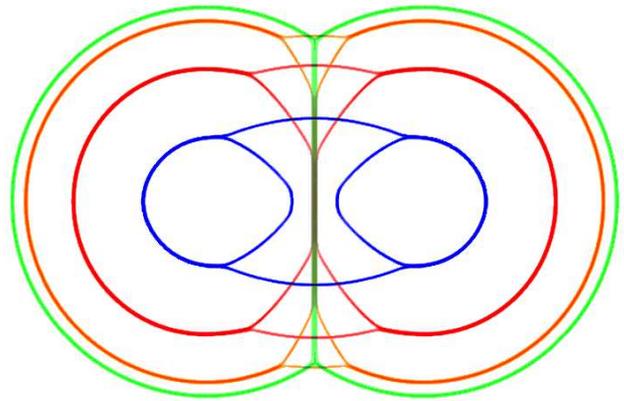}}
\caption{Results from a field theoretic simulation for $(1,1)+(1,-1)\rightarrow(2,0)$ with $\kappa=0.8$, showing the decomposition of the initial Y-junctions. Snapshots are shown for times $t=0$, $0.36$, $0.8$ and $1.24$; each representing the cummulative projection of energy density on the plane. The simulation had lattice spacing $\Delta x=0.5/\eta$, with $\eta=90$.}
\label{fig:decomp}
\end{figure}

An interesting outcome emerges in the field theoretic
case of a $(1,1)+(1,-1)\rightarrow(2,0)$ junction in that the
Y-junction itself can decompose as illustrated in
Fig.~\ref{fig:setup2}. Whether or not
this occurs will 
depend upon the tensions of the $(2,0)$, $(1,1)$ and $(1,0)$ strings
(which cover all tensions involved due to the symmetry in our
parameter choice) and upon the orientation of the external strings. By
coincidence $R=\mu_{0}/2\mu_{1}$ is approximately constant
for $(1,1)+(1,-1)\rightarrow(2,0)$ over the range of $\kappa$ that we consider here ($0.8 \le \kappa \le 0.95$). As a result $R$ is effectively $0.56$ over this entire range, essentially fixing both the initial orientations and the large-scale dynamics. However, the ratio of $\mathcal{R}=\mu_{1}/2\mu_{2}$ (where $\mu_2$ is the tension of a (1,0) or (0,1) string) varies from $0.86$ to $0.78$ as $\kappa$ is increased over that range. Of course $\mu_{0}/2\mu_{2}$ must decrease
accordingly since $R$ is almost constant, and hence both types of composite string become increasingly stable against this decomposition. This can be seen in our results for $\kappa=0.8$, shown in 
Fig.~\ref{fig:decomp}, and those for $\kappa=0.95$ which have already been presented in
Fig.~\ref{fig:results95}, where in the new case the junctions
decompose and in the previous case they remain stable. 

The final state of the system is no longer two loops for the cases
when the Y-junctions decompose and instead the Y-junctions meet such that the bridge has effectively been cut along
its length. There are then two separate loops in $\phi$ which sit at
opposite sides of an elongated $\psi$ loop to which they are each
bound. These inner $\phi$ loops then collapse before the outer $\psi$
loop. 

\begin{figure}
\resizebox{\columnwidth}{!}{\includegraphics{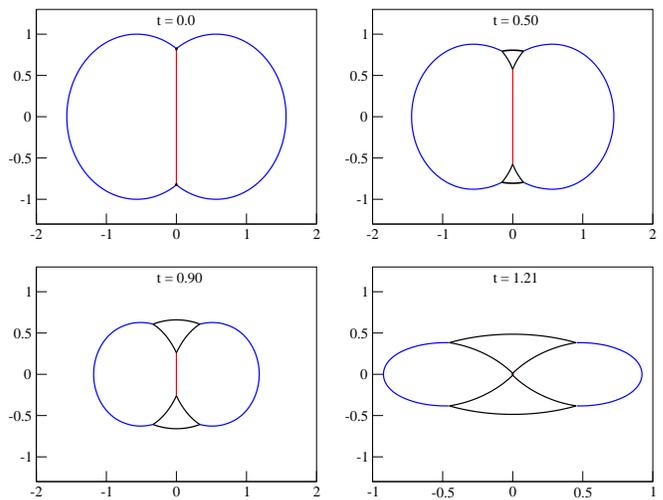}}
\caption{The perturbed butterfly loop corresponding to the
  $\kappa=0.8$ field theory case of Fig.~\ref{fig:decomp},
using a perturbation parameter $h=0.01$ - the instability grows and the loop
  is unstable.} 
\label{fig:pert_but_1}
\end{figure}

Remarkably, we can also analyse this instability of the Y-junctions to
decomposition in the Nambu-Goto picture, by changing our initial
conditions to start with the decomposed state, but with the three
Y-junctions very close together. The internal strings between these
junctions are taken to be circular arcs since this approximately
represents the curving effects of the tensions that are applied to the
ends of these strings. These are the initial conditions shown for time
$t=0$ in Fig.~\ref{fig:pert_but_1} although the junctions are so close
that they cannot be resolved in this plot. However the snapshot for
$t=0.50$ shows that in this case, for which the tensions correspond to
the $\kappa=0.8$ field theory values, the three Y-junctions are
further separated and approximately circular arcs have grown between, which are now readily
apparent. That is, this and later times indicate that the Nambu-Goto
evolution also yields the growth of the internal strings and so matches the field theoretic results. However, applying this
procedure to the $\kappa=0.95$ case, shows that the
$(1,1)+(1,-1)\rightarrow(2,0)$ junction is then stable to collapse,
again agreeing with the field theory results, as is clear from Fig.~\ref{fig:results95}. The precise numerical results do differ for the unstable case however, since in the field
theoretic case there is a considerable role of the small-scale
microphysics in the breakup of the junction and the breakup happens
very much faster than is seen in the Nambu-Goto case. Additionally,
the use of circular arcs here is artificial and would not be expected
to be wholly accurate. However, the approximate matching of the
critical value for $\kappa$ at which the breakup occurs can be
considered a considerable success for the less computationally
intensive CKS approach, and crucially, as we show below, it allows us to make a prediction based purely on the Nambu-Goto results as to when a junction will and will not be stable  to decomposition into many junctions. 

\begin{figure}
\resizebox{\columnwidth}{!}{\includegraphics{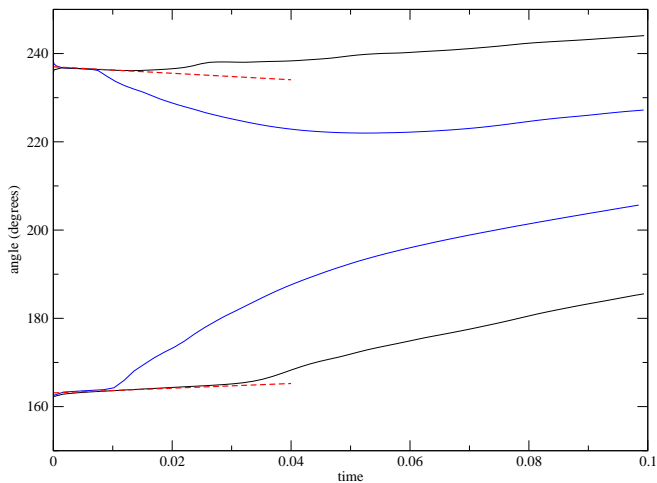}}
\caption{Numerical (solid lines) and analytic (dashed red lines) evolution
  of the angle $\varphi$, for junction A1, for two cases with $\rho>0$
  and two with $\rho<0$; all close to the critical value $\rho = 
  0$ (blue lines are closer to the critical value). For
  $\rho <0$ (bottom curves) the Y-junction is said to be 
  unstable, since vertex A moves along the butterfly wing until it
  reaches 180$^{\circ}$, and then it starts moving 
  towards the centre of the big arc, as shown in
  Fig.~\ref{fig:pert_but_1}. For $\rho >0$ (top curves),  
  vertex A moves downwards, leading to a stable
  Y-junction. The analytic approximations are calculated using
  $R=0.561$ and equations (\ref{eq:rholesszero}) and
  (\ref{eq:rhobigzero}), which  
  are linear truncations (in time), and only hold for small times since
  $\ell^{A_1}_{\rm min}(t)\sim h$. We choose $h=0.01$.}   
\label{fig:vertex}
\end{figure}

Using the Nambu-Goto approach, the stability of Y-junctions can be
studied analytically, at least for small times. In appendix
\ref{sec:appendixC}, we describe this procedure in detail for junction $A_1$ 
in Fig.~\ref{fig:setup2}. We find that the
initial perturbation in the decomposed state grows or collapses
according to the angle $\rho$ (see Fig.~\ref{fig:setup2}), which
purely depends on the original 
butterfly tensions, the tension of the
small arc segments, $\mu_2$, and the size of the perturbation
$h$, in the following way 
\be\label{rhoeq}
\rho=\frac{\pi}{2}-\cos^{-1}\left(\mathcal{R}\right)-\cos^{-1}\left({R}-h\right),
\ee
where $\mathcal{R} =\frac{\mu_1}{2\mu_2}$ and, as before, $R =
\frac{\mu_0}{2\mu_1}$. 
Therefore, for a given pair of tensions $\mu_0$ and $\mu_1$ there is a 
critical tension $\mu_2 = \mu_{crit}$ (for a small fixed $h$), for
which $\rho = 0$. 
Below and above this critical limit, there are two distinctive
behaviours: one in which the perturbation grows (as in
Fig.~\ref{fig:pert_but_1}) and one in which it either does not grow
significatetly or shrinks, and simply resembles the original 
butterfly (as in the field theory simulation of
Fig.~\ref{fig:results95}). To illustrate this, we can use the analytic
results for $\dot{s}_i$ and the evolution of the vertex
$\mathbf{X}_{A_1}$ as a function of the right (and/or left movers), which
explicitly is given by (see Appendix \ref{sec:appendixC}) 
\be
\dot{\mathbf{X}}_{A_1} = -\frac{1}{\sum_j\mu_j}\sum_j \mu_j(1-\dot{s}_j)\bb'_j.
\ee
To map the junction movement in real space, it is useful to define
the angle 
\be
\tan(\varphi) = \left(\frac{\dot{X}^{A_1}_y}{\dot{X}^{A_1}_x}\right),
\ee
which shows the direction with respect to the $x$-axis for $t>0$. The
explicit treatment of this angle is given in appendix 
\ref{sec:appendixC}. The main result is that there exists a
discontinuity in $\varphi$ when going from $\rho>0$ to $\rho<0$, as
Fig.~\ref{fig:vertex} shows\footnote{Note that a negative $\rho$
  corresponds to the arc $A_1A_2$ in Fig.~\ref{fig:setup2} going from
  concave, as shown, to convex. Equivalently, the centre of the circle
  from which the arc is formed moves from below the arc, as in the
  figure, to above it.}. This shows how the evolution of 
the splitting of the Y-junction in the original butterfly depends
mainly on the {\it{initial local curvature}} of the strings involved.   

When $\rho>0$ (see Fig.~\ref{fig:setup2}), strings $2$ and $3$ are
\textquotedblleft competing\textquotedblright in $\sigma$-space while
the butterfly wing (string $1$) is not contributing much (note that
for small times $\dot{s}_1 = 0$ to first order in $t$). After some
time and in real space, the vertex $A_1$ moves downwards with
an initial angle of $\varphi\geq \pi+\tan^{-1}(\sqrt{1-R^2}/R)$ (with
the equality in the limit of $\rho\rightarrow 0$) from the
$x$-axis, as can be seen in Fig.~\ref{fig:vertex} (see Appendix
\ref{sec:appendixC} for details of the calculation). In this case the
perturbation does not grow and, for a tension $\mu_2$ big enough, it may even
collapse faster than the central bridge does. However, for $\rho<0$
the local curvature is such that the strings of the
triangular perturbation grow in $\sigma$-space. In real space, vertex $A_1$
initially moves rapidly away from the $y$-axis and almost along the
butterfly wing (string 1), which corresponds to an initial angle of
$\varphi\leq\pi-\tan^{-1}(R/(1-\sqrt{1-R^2})$ from the $x$-axis (see 
Appendix \ref{sec:appendixC} for details). In figures \ref{fig:pert_but_1} and
\ref{fig:vertex}, one can see this initial evolution. Later in the
evolution (when the angle $\varphi$ reaches $\pi$), the segment $A_1A_2$
changes from convex to concave, and the vertex $A_1$ evolves
like any other point on the big arc segment, hence moving towards the
centre of the butterfly wing, as one can see in the last two plots of
Fig.~\ref{fig:pert_but_1}. Therefore, for $\rho<0$ the perturbation
grows for some time (which depends on how negative $\rho$ initially
is), implying the original butterfly Y-junction is unstable, leading
to the criterion for stability based on simply obtaining the value for
$\rho$. 

\section{Discussion} \label{sec:discuss}
The role of extended objects in the evolution of the Universe has been studied for many years owing to their intrinsically interesting properties, their crucial role in string/M theory, repeated appearance in GUT models and their presence in many condensed matter systems. Here we have extended previous studies on the dynamics of cosmic strings to include the situations envisioned in models of cosmic superstrings, namely the existence of bound-state strings, and the junctions that join them together\cite{Copeland:2003bj}. While analytic studies of the superstrings themselves have met with limited success \cite{Jackson:2004zg,Jackson:2007hn}, the framework for numerical studies of field theory defects is well-developed, and simple models of bound-state strings \cite{Saffin:2005cs} can be used to understand the large-scale properties of networks \cite{Copeland:2005cy,Hindmarsh:2006qn,Rajantie:2007hp,Urrestilla:2007yw}, as well as more detailed analyses on their individual collisions \cite{Bevis:2008hg,Salmi:2007ah}. 

The great utility of Nambu-Goto dynamics is the huge reduction in the degrees of freedom compared to field theory simulations, allowing for numerical computations with far larger dynamic range. However, as is well known, the Nambu-Goto dynamics break down as a description of gauged cosmic strings when two vortices cross, loops contract to a point, or when junctions collide. This is well established for the case of the usual Abelian cosmic strings, and algorithms have been developed (motivated by the field theory results) which nevertheless allow the Nambu-Goto equations to be consistently used to model the evolution of a network of cosmic strings  (see \cite{Vilenkin:2000} for details). This same procedure has not yet been established for the case of strings with junctions, and has been a key goal of this paper. By comparing the modified Nambu-Goto equations for such strings, with the field theory evolution for similar strings we have been able to establish how well they follow each other, 
and where differences emerge. 
By concentrating on a simple loop configuration, that nevertheless encapsulates important dynamics, we have been able to explore various properties and directly probe the relationship between these two approaches, answering questions such as: What happens when junctions collide? What is the effect of loop-collapse on the wings of the butterfly loop? What can we say about the instability of junctions as the bound states try to unzip\cite{Bettencourt:1994kc}?

In terms of the general dynamics, we have seen that the evolution of strings with junctions are remarkably well modelled by the Nambu-Goto action \cite{Copeland:2006eh,Copeland:2006if,Copeland:2007nv}, despite the curvature at the junctions being high. However, the field theory approach has opened up a crucial new instability that is not present in the usual Nambu Goto formalism, namely the break up of a  junction into three new junctions and the corresponding unzipping of the composite string. In other words we still need the field theory to understand the outcome of collisions of strings and collisions of junctions. 

Examining the field theory of strings with different binding energies it became clear that for weakly-bound composites the junctions could cause the strings to unzip, care of the new instability. Once this was realized it became possible to model this within the Nambu-Goto dynamics by introducing the potentially unstable triangular configuration at the junctions, and solving for its evolution. Remarkably, this then allowed us to predict in terms of the angle $\rho$ (or in terms of the string tensions) when a junction would be unstable to this decay mode, and when it would remain stable, and the results agreed well with the field theory simulations. With this done there is again excellent agreement between the Nambu-Goto simulations and the field theory. The key factor here is that given this understanding of the instability, it is once again possible to perform,  with confidence, large-scale cosmological simulations of cosmic superstrings using these modified Nambu-Goto strings, in the same manner that it is possible to perform simulations of ordinary cosmic strings using Nambu Goto equations.  

\section{Acknowledgments} 
We acknowledge financial support from the Royal society (EJC), STFC (NB and GN) and the University of Nottingham (AP). Field theory simulations were performed on the UK National Cosmology Supercomputer (COSMOS) supported by SGI, Intel, HEFCE and STFC. 
We would also like to thank Tom Kibble for many useful discussions.
\appendix

\section{Equations of motion} \label{sec:appendixA}

Working in the conformal gauge provides an elegant way to derive the energy conservation equation at each junction.
Consider the action given in (\ref{action1}) where we now introduce the parameter $\delta_i$ related to the orientation of the strings at the junction:
 \bea 
 S&=&-\sum_i\mu_i \delta_i \int \d\tau \int \d\si_i\, \big[\Theta(s^{B_i}_i(\tau)-\si_i)\Theta(-s^{A_i}_i(\tau)+\si_i) 
 \nonumber
 \\
 && \qquad \qquad \qquad \qquad \qquad
 \times \sqrt{-x_i'^2 \, \dot x_{i}^{2} } \big]
 \nonumber\\
 &&+\sum_J\sum_i\int \d \tau \,
 f^J_{i \mu}  \cdot [x_{i}^\mu( \tau,s^J_i (\tau) )-{X}_J^\mu(\tau)], 
 \label{action}
 \eea
where $A_i$($B_i$) is the junction where $\sigma_i$ starts(ends).

Varying the action (\ref{action}) with respect to $X^\mu_J$ we have
\be
\sum_i f^J_{i \mu}=0.
\label{summ}
\ee
The boundary conditions, on the other hand, are
$
\delta_i\mu_i \,  ({ x_{i}^{\mu} }' + \dot s_{i,J} \dot x_{i}^{\mu}) = f_{i,J}^{\mu}
$
(evaluated at each junction) which, using Eqs.~(\ref{eq:wave}) and (\ref{summ}) gives 
\bea
0 &=&  \sum_{i}  \mu_i \delta_i \,  \left[{ x_{i}'^{\mu} }(u^J_i) + \dot s_{i}^J \dot x_{i}^{\mu}(v^J_i)\right]  
\label{Xeqn}
\\ \nn
&=& \sum_i\mu_i \delta_i  \,  \left[ ( \dot s_{i} + 1)
a_{i}'^{\mu}(u^J_i)  +  ( 1- \dot s_{i} ) b_{i}'^{\mu}(v^J_i)
\right]  
\eea
where
\be
u_i^{J}(\tau) = s^{J}_i(\tau)+ \tau \, , \qquad  v_i^{J}(\tau) = s^{J}_i(\tau)- \tau.
 \ee
Note Eq.~(\ref{Xeqn}) reproduces Eq.~(\ref{vec-sum}) for the case of a static string $\dot s_{i}=0$. On imposing the temporal gauge $\tau = t = x_i^0(\tau,\sigma_i)$, the $\mu=0$ component of this equation reduces to energy conservation at each junction:
\begin{equation}
 \delta_1\mu_1\dot{s}^J_1+\delta_2\mu_2\dot{s}^J_2+\delta_3\mu_3\dot{s}^J_3=0.
\end{equation}

Now, the constraint that the three strings meet at junction $B$ leads to
\bea
2 \dot{{X}}^\mu_B =  ( \dot s^B_{i} + 1) a_{i}'^{\mu}(u^B_i) 
+  ( \dot s_{i} - 1 )  b_{i}'^{\mu}(v^B_i)  .
\label{22}
\eea
Then we proceed depending on the relative orientation between the $3$ strings in question. To illustrate our method, consider the specific example where $\delta_1 = \delta_2 = 1$, $\delta_3 = -1$ for junction $B$. That is, the incoming waves at junction $B$ are $b_{1}'$, $b_{2}'$ and $a_{3}'$.
Eliminating the outgoing waves  $a_{1}'$, $a_{2}'$ and $b_{3}'$ between (\ref{Xeqn}) and (\ref{22}) gives
\bea
(\sum_j \mu_j)\dot{{X}}^\mu_B = \mu_3(1+\dot{s}^B_3)a_{3}'^{\mu}- \sum_{i=1,2}{\mu}_i b_{i}'^{\mu} (1-\dot{s}^B_i).
\label{33}
\eea
Eliminating $\dot{{X}}^\mu_B$ from Eqs.~(\ref{22}) and (\ref{33}) we can solve for the unknown outgoing waves. This gives 
\bea
-\mu(1+\dot{s}_1^B)\ba'_1 = 2\mu_2(1-\dot{s}_2^B)\bb'_2-2\mu_3(1+\dot{s}_3^B)\ba'_3 \nonumber \\ 
+\Big(2\mu_1-\sum_j\mu_j\Big)(1-\dot{s}_1^B)\bb'_1 
\label{eq:thusgen}
\eea
and two similar equations for $\ba'_2$ and $\bb'_3$. Squaring these equations and using the gauge conditions (\ref{confg}) we find eq.~(\ref{sdotgen})
with $\delta_1 = \delta_2 = 1$ and $\delta_3 = -1$.

\section{Analytic result for the butterfly configuration with two vertices} \label{sec:appendixB}

As discussed in section \ref{sec:nambugotoA}, analytical progress can be readily made for the period before the two junctions become causally connected. However, since the shortest communication route is the central bridge string, and the symmetry of the butterfly configuration ensures that the only dynamics of the central bridge is for it to change it length, then simple analytical results can be obtained for later times also, as explained below. 

The initial conditions are a string lying on the $y$-axis (string 0)
and two arcs of unit circles (strings 1 and 2) in the $x-y$
plane. If is useful to introduce an angle $\gamma$ where, for a static initial configuration, $\cos\gamma=-R$. We then have: 
\begin{eqnarray}
\label{butter-config} 
\bx_0(t=0,\si_0)\eq(0,\si_0,0),\qquad|\si_0|<\sin\gamma,\\
\bx_{1}(t=0,\si_1)\eq\left(-\cos\gamma+\cos\si_1,
\sin\si_1,0\right), \; \; \; |\si_1|<\gamma,\nonumber \\
\bx_{2}(t=0,\si_2)\eq\left(\cos\gamma-\cos\si_2, \nonumber
\sin\si_2,0\right), \; \; \; |\si_2|<\gamma.
\end{eqnarray}
If we label the lower vertex as $A$ and the upper one as $B$, then $\si_i$ increases towards
junction $B$ for all strings. Because of symmetry it is sufficient to study one junction, say $B$. Therefore, at $t=0$
\begin{eqnarray}
\vect{a}'_0=\mathbf{b'}_0 &=& (0,1,0), \nonumber\\
\vect{a}'_1=\mathbf{b'}_1 &=& (-\sin\si_1,\cos\si_1,0),\nonumber\\
\vect{a}'_2=\mathbf{b'}_2 &=& (\sin\si_2,\cos\si_2,0).
\end{eqnarray}

Since the string 0 is simply stationary and on the $y$ axis for all
times, then the above $t=0$ result is valid for all $t$ and the
ordinarily difficult to handle emitted waves are simply the waves set
by the initial conditions. When the junctions do come into casual
contact via string 0, the situation is completely unchanged and hence
simple analytical results can still be obtained. Furthermore, string 0
is the shortest communication route, while the results below can be
used to show that the two longer routes are always too long to become relevant 
before the central bridge has collapsed. 

Conservation of energy (\ref{eq:energycks}) implies that $R \dot{s}_{0}^B=-\dot{s}_{1}^B$ and hence after integration we have:
\begin{equation}
\label{eqn:s0funcs1}
s_0^B(t) = \sin\gamma - \frac{1}{R} \left( s_1^B(t) - \gamma \right).
\end{equation}
Thus, it is sufficient to determine the function $s_{1}^B(t)$. In this case we have 
$c_{1}=c_{2}=\cos(s_{1}^B-t)$ and $c_{0}=2\cos^{2}(s_{1}^B-t)-1$ and
so it is useful to denote $\Delta=t-s_{1}^B$, whose derivative is
$\dot{\Delta}=1-\dot{s}_{1}^B$. Therefore Eq. (\ref{sdot_b}) implies: 
\begin{equation} 
\dot{\Delta}=\frac{1-R^2}{1+R\cos\Delta}. 
\end{equation}
This simple separable equation can be integrated to give:
\begin{equation}
t \sin^{2}\gamma = -\cos \gamma \sin\Delta + \Delta - \cos \gamma \sin \gamma + \gamma.
\end{equation}
Together with the definition of $\Delta$ and Eq. \ref{eqn:s0funcs1}, we
now have $t$, $s_{0}^B$ and $s_{1}^B$ specified as functions of the
variable $\Delta$. 

\section{Stability of $Y$-junctions using the Nambu-Goto approximation} \label{sec:appendixC}

Using the same small time analysis as in Appendix
\ref{sec:appendixB}, we can study the stability of the initial
perturbation introduced in the butterfly configuration. 
We will consider junction $A_1$ (see
Fig.~\ref{fig:setup2}), however this analysis can be equally
applied to any other junction. The angles shown in
Fig.~\ref{fig:setup2} are completely determined by the initial
equilibrium conditions and hence they are defined in terms of the
tensions $\mu_1$ and $\mu_2$. The position of junction $A_1$ in $\sigma$-space is 
\bea
s^{A_1}_1(0) &=& \gamma = \pi - \cos^{-1}\left(\frac{\mu_0}{2\mu_1}-h\right),\nonumber \\
s^{A_1}_2(0) &=&\eta = \cos^{-1}\left(\frac{\mu_0}{2\mu_2}\right) - \alpha, \nonumber \\
s^{A_1}_3(0) &=& \rho = \gamma - \alpha - \frac{\pi}{2}.
\eea 
where $\alpha = \cos^{-1}\frac{\mu_1}{2\mu_2}$ and $h$ is the distance between junction $A_1$ (or $A_2$) and junction $A_3$ in Fig.~\ref{fig:setup2}.

We will now analytically demonstrate that the behaviour of the
perturbation (i.e. whether it grows or collapses) depends on whether
the angle $\rho$ is positive or negative. For that we will consider
both cases and study the behaviour of $\dot{s}_i^{A_1}$ and
$\mathbf{\dot{X}}^{A_1}$. We will drop out the junction index ($A_1$)
from now on for simplicity.\\  

{ \bf{Case I: $\rho < 0$}}\\

The initial configuration comprises of three strings with tensions $\mu_1$, $\mu_2$ and
\bea
\mathbf{b'}_1(t=0, \sigma_1) &=& (\sin{\sigma_1},\cos{\sigma_1},0), \nonumber \\  
\mathbf{b'}_2(t=0, \sigma_2) &=& (-\sin{\sigma_2},\cos{\sigma_2},0),\nonumber \\ 
\mathbf{b'}_3(t=0, \sigma_3) &=& (-\cos{\sigma_3},\sin{\sigma_3},0). 
\eea

At a later time $t$ the incoming waves at junction $A$ are
\bea
\mathbf{b'}_1(t,s_1(t)) &=& (\sin{(s_1(t) - t)},\cos{(s_1(t) - t)},0), \nonumber \\
\mathbf{b'}_2(t,s_2(t)) &=& (-\sin{(s_2(t) - t)},\cos{(s_2(t) - t)},0), \nonumber \\
\mathbf{b'}_3(t,s_3(t)) &=& (-\cos{(s_3(t) - t)},\sin{(s_3(t) - t)},0). \nonumber \\
\eea
Taylor expanding $s_i$ we get  $s_i(t) = s_i(0) + \lambda_i t^2 + ...$
(remember $\dot{s}_i = 0$ initially), and using the relations between
the angles we find (to first order in $t$) 
\bea
c_{1} &=& \cos{(2\alpha + 2t)},\nonumber \\
c_{2} &=& -\cos{\alpha}, \nonumber \\
c_{3} &=& -\cos{(\alpha + 2t)}.
\eea
Now, using equation (\ref{sdot}), linearising in $t$ and defining
$\mathcal{R}=\cos{\alpha}=\frac{\mu_1}{2\mu_2}$ (which is always less than unity due to the
triangle inequalities) we find  
\beq
\dot{s}_{1} = -\dot{s}_{3} =-\frac{1}{\sqrt{1-\mathcal{R}^2}}t,\qquad
\dot{s}_{2} = \left(\frac{2\mathcal{R}-1}{1-\mathcal{R}^2}\right)t. 
\eeq \\
{ \bf{Case II: $\rho > 0$}}\\

In this case, the initials conditions can be written as 
\bea
\mathbf{b'}_1(t=0,\sigma_1) &=& (\sin{\sigma_1},\cos{\sigma_1},0), \nonumber \\  
\mathbf{b'}_2(t=0,\sigma_2) &=& (-\sin{\sigma_2},\cos{\sigma_2},0),\nonumber \\ 
\mathbf{b'}_3(t=0,\sigma_3) &=& (-\cos{\sigma_3},-\sin{\sigma_3},0). 
\eea
Following the same procedure we find
\beq
\dot{s}_{1} = 0,\qquad
\dot{s}_{2} = -\dot{s}_{3}=-\frac{2\sqrt{1-\mathcal{R}^2}}{1+\mathcal{R}}t.
\eeq \\
Having the analytic expressions for $\dot{s}_i$ (which of course can
be extended further than first order in $t$) we can easily find the
analytical expression for $\dot{\mathbf{X}}$ using equations
(\ref{Xeqn}) and (\ref{22}), to obtain the expression 
\be
\dot{\mathbf{X}} = -\frac{1}{\sum_j\mu_j}\sum_j \mu_j(1-\dot{s}_j)\bb'_j.
\label{eq:appvertex}
\ee
In order to study the motion of the vertex in real space, we define the angle 
\be
\tan(\varphi) = \left( \frac{\dot{X}_y}{\dot{X}_x}\right). 
\ee 
Since the expression for this angle is very complicated, one can take
the limit in which the perturbation size $h$ tends to zero, and also
consider small deviations from the $\rho=0$ case, either with positive
or negative $\rho$. The critical tension $\mu_2$ which leads to
$\rho=0$ is obtained by setting (\ref{rhoeq}) to zero and solving for $\mu_2$,
resulting in
\beq
\mu_{crit}=\frac{\mu_1}{2\cos\left(\cos^{-1}(R-h)-\pi/2\right)}, 
\eeq
Therefore, in the limit $h \rightarrow 0$ and $\mu_2 = \mu_{crit}$
equation~(\ref{eq:appvertex}) reduces to
\be
\frac{\dot{X}_y}{\dot{X}_x} = -\frac{R}{1+\sqrt{1-R^2}} + \frac{-1+3R^2+\sqrt{1-R^2}}{R^2(1+\sqrt{1-R^2}-2R^2\sqrt{1-R^2})}t
\label{eq:rholesszero}
\ee 
for $\rho < 0$, and
\be
\frac{\dot{X}_y}{\dot{X}_x} = \frac{\sqrt{1-R^2}}{R} - \frac{2+R^2-2\sqrt{1-R^2}}{2R^4}t 
\label{eq:rhobigzero}
\ee
for $\rho > 0$. Notice that, as one should expect, in the critical tension limit
$\mathcal{R}$ drops out from the expressions, and only
$R=\frac{\mu_0}{2\mu_1}$ appears. For both cases ($\rho>0$ and $\rho<0$),
$\dot{X}_x$ is initially positive, so it is the $y$ direction which
changes. For $\rho>0$, the vertex $A_1$ moves with an initial angle of
\be
\varphi=\pi+\tan^{-1}\left(\frac{\sqrt{1-R^2}}{R}\right)
\ee
in the critical limit ($\mu_2=\mu_{crit}$), and bigger angles for
$\mu_2>\mu_{crit}$; therefore the perturbation does not
grow. In contrast, for $\rho<0$, the vertex $A_1$ moves away from the
$y$-axis, with an initial angle of 
\be
\varphi=\pi-\tan^{-1}\left(\frac{R}{1+\sqrt{1-R^2}}\right),
\ee
which is practically along the butterfly wing. In this case, the junctions
separate initially from each other and the butterfly configuration is
unstable.

\end{document}